\documentclass[twocolumn,amsmath,floatfix]{revtex4}
\usepackage{latexsym}
\usepackage{amssymb}
\usepackage{pdfpages}
\usepackage{graphics}

\begin{document}
\newcommand{\ja}{Jakubassa-Amundsen}

\newcommand{\bfx}{\mbox{\boldmath $x$}}
\newcommand{\bfq}{\mbox{\boldmath $q$}}
\newcommand{\bfnabla}{\mbox{\boldmath $\nabla$}}
\newcommand{\bfsigma}{\mbox{\boldmath $\sigma$}}
\newcommand{\bfSigma}{\mbox{\boldmath $\Sigma$}}
\newcommand{\bfsigmas}{\mbox{{\scriptsize \boldmath $\sigma$}}}
\newcommand{\bfeps}{\mbox{\boldmath $\epsilon$}}
\newcommand{\bfGamma}{\mbox{\boldmath $\Gamma$}}
\newcommand{\bfalpha}{\mbox{\boldmath $\alpha$}}
\newcommand{\bfA}{\mbox{\boldmath $A$}}
\newcommand{\bfP}{\mbox{\boldmath $P$}}
\newcommand{\bfF}{\mbox{\boldmath $F$}}
\newcommand{\bfe}{\mbox{\boldmath $e$}}
\newcommand{\bfd}{\mbox{\boldmath $d$}}
\newcommand{\bfes}{\mbox{{\scriptsize \boldmath $e$}}}
\newcommand{\bfn}{\mbox{\boldmath $n$}}
\newcommand{\bfW}{{\mbox{\boldmath $W$}_{\!\!rad}}}
\newcommand{\bfM}{\mbox{\boldmath $M$}}
\newcommand{\bfK}{\mbox{\boldmath $K$}}
\newcommand{\bfI}{\mbox{\boldmath $I$}}
\newcommand{\bfQ}{\mbox{\boldmath $Q$}}
\newcommand{\bfp}{\mbox{\boldmath $p$}}
\newcommand{\bfk}{\mbox{\boldmath $k$}}
\newcommand{\bfks}{\mbox{{\scriptsize \boldmath $k$}}}
\newcommand{\bfs}{\mbox{\boldmath $s$}_0}
\newcommand{\bfv}{\mbox{\boldmath $v$}}
\newcommand{\bfw}{\mbox{\boldmath $w$}}
\newcommand{\bfb}{\mbox{\boldmath $b$}}
\newcommand{\bfxi}{\mbox{\boldmath $\xi$}}
\newcommand{\bfzeta}{\mbox{\boldmath $\zeta$}}
\newcommand{\bfr}{\mbox{\boldmath $r$}}
\newcommand{\bfrs}{\mbox{{\scriptsize \boldmath $r$}}}

\renewcommand{\theequation}{\arabic{section}.\arabic{equation}}

\title{\Large\bf Numerical test of polarization sum rules for the triply differential bremsstrahlung cross section in electron-nucleus encounters}

\author{D.~H.~Jakubassa-Amundsen\\
Mathematics Institute, University of Munich, Theresienstrasse 39,\\ 80333 Munich, Germany}

\date{\today}


\vspace{1cm}

\begin{abstract}  
Inspired by the work of Pratt and coworkers on a sum rule for the polarization correlations in electron bremsstrahlung when the outgoing
electron is not observed, we derive the corresponding sum rule for the elementary process of bremsstrahlung.
This sum rule is valid for arbitrary electron wavefunctions provided the electron is emitted in the reaction plane.
The numerical evaluation of this sum rule within the Dirac partial-wave theory for bare inert spin-zero nuclei and collision  energies in the range of
$1-10$ MeV reveals violations for high nuclear charge. Such violations serve as a measure of the inaccuracies in the bremsstrahlung calculations.
\end{abstract}

\maketitle


\vspace{0.5cm}
\section{Introduction}
The elementary $(e,e'\gamma)$ process of bremsstrahlung in electron-atom or electron-nucleus collisions has been thoroughly studied in the past \cite{OM,Na,HN}.
The measurement of polarization correlations between the incoming electron and the outgoing photon provides
a sensitive test of the relativistic dynamics of polarized electrons in strong nuclear fields.
Profiting from recent advances in polarimetry \cite{Ta11b,Ba10,Ta13}, the measurement of polarization correlations has become a field of current interest,
covering collision energies up to 3.5 MeV \cite{Ba10,Ta15,Ni}.
Also in low-energy nuclear structure studies experiments with polarized electrons are planned, where knowledge on bremsstrahlung is crucial since it contributes essentially to the background of excitation spectra
\cite{MI1,GP1,JK}
or to photon angular distributions from the coincident nuclear excitation and decay \cite{PW} in case of heavy nuclei \cite{JP}.

Sum rules for the polarization correlations have always been a matter of interest. A sum rule pertaining to elastic electron (potential)  scattering,
$L^2+R^2+S^2=1$, which involves the three correlation parameters $L,R$ and $S$ (relating to the initial electron spin polarization along the three coordinate axes),
has been known for a long time \cite{Mo}.
The equivalence between an elastically scattered electron and a bremsstrahlung photon at the short-wavelength limit,
at collision energies high enough such that the electron's rest mass can be neglected, has therefore led to an approximate sum rule for bremsstrahlung in the case of strong nuclear fields \cite{Jaku12},
\begin{equation}\label{1.1}
C_{32}^2+C_{12}^2+C_{20}^2\;\approx\;1,
\end{equation}
where $C_{32},C_{12}$ and  $C_{20}$ describe the polarization correlations between the incoming electron (with spin polarization 
along the coordinate axes as above) and a circularly polarized photon,
assuming that the scattered electron is not observed.

Most recently, Pratt and coworkers have succeeded in deriving an exact sum rule which involves all seven  polarization correlations (between the incoming electron and a linearly or circularly polarized photon),
which are allowed by the time-reversal invariance of the transition matrix element \cite{PRT,PMS}.
This sum rule,
\begin{equation}\label{1.2}
C_{32}^2+C_{12}^2+C_{20}^2+C_{03}^2+C_{31}^2+C_{11}^2-C_{23}^2\;=\;1,
\end{equation}
holds independently of collision energy, nuclear charge number and photon momentum \cite{PMS}.
The only condition for its validity is the restriction to a  single partial wave in the final electronic state, as well as to only two
of its  magnetic substates
(those of opposite sign).
A corresponding sum rule was also derived for photoionization, which can be considered as the reverse process of bremsstrahlung
 near the short-wavelength limit, and which is easier to handle theoretically \cite{PRT,PMS,PL,SP}.

There is yet another sum rule, also put forth by
Pratt and coworkers \cite{PMS},
\begin{equation}\label{1.3}
C_{32}^2+C_{12}^2+C_{03}^2\;\approx \;1,
\end{equation}
which is approximately valid at small nuclear charge numbers  and high collision energies.

In the present work a sum rule of the type (\ref{1.2}) is derived for the polarization correlations occuring in the elementary process of bremsstrahlung 
(where outgoing electron and photon are observed in coincidence).
In contrast to the doubly differential case studied in  \cite{PMS} where the final electronic partial waves add {\it incoherently}, the triply differential case considered below involves a {\it coherent} sum over these partial waves. This is the reason why the corresponding sum rule is exact (for coplanar geometry) without any restriction
on the number of final partial waves.

Early calculations of the bremsstrahlung process, going beyond the plane-wave Born approximation \cite{BH}, have applied semirelativistic
(Sommerfeld-Maue) wavefunctions for the electronic scattering states \cite{OM,HN}. However, for heavy nuclei, this approach is inferior to the
 Dirac partial-wave (DW) theory which uses exact solutions to the Dirac equation and which has become standard nowadays \cite{TP71,TP73,YS,MYS}.
Since for very high collision energies the DW theory suffers from serious convergence problems, its applicability is restricted   to  energies below 30 MeV \cite{Jaku16}.

The paper is organized as follows.
In Section II the partial-wave theory for bremsstrahlung is outlined and the sum rule is explicitly derived,
profiting from a symmetry relation for the transition amplitudes.
Section III provides a numerical test of the sum rules. The  charge numbers $Z_T=4$ and 82 at collision energies 1 MeV and 10 MeV, respectively, are considered.
Concluding remarks are given in Section IV.
Atomic units ($\hbar=m=e=1)$ are used unless indicated otherwise.

\section{Bremsstrahlung theory and the polarization correlations}
\setcounter{equation}{0}

In this section an evaluation of the spin-dependent triply differential bremsstrahlung cross section  within the DW theory 
is given, from which the polarization correlations are obtained. Finally a proof of the sum rule
\begin{equation}\label{2.1}
C_{320}^2+C_{120}^2+A^2+P_1^2+P_2(0)^2+P_2(90^\circ)^2-C_{230}^2\;=1,
\end{equation}
which is the equivalent of the sum rule (\ref{1.2})
for the ele\-mentary process of bremsstrahlung, is provided.
For the definitions see subsection B.
Note the identification $A \equiv C_{200},\;P_1 \equiv C_{030},\; P_2(0)\equiv -C_{310}$ and $P_2(90^\circ)\equiv C_{110}$.

\subsection{Evaluation of the cross section}

We describe the initial spin polarization $\bfzeta_i$ of the electron in terms of the coefficients $a_{m_i}\;\;(m_i=\pm \frac12)$ of the up and down spinors $\chi_{1/2} = {1 \choose 0}$ and $\chi_{-1/2} = {0 \choose 1}$ \cite{Ros}.
Furthermore, the photon polarization $\bfeps_\lambda$ is represented as a linear combination
$\bfeps_\lambda = \sum_{\sigma =\pm} f_\sigma^\ast  \bfeps_\sigma$ of the basis vectors $\bfeps_+$ and $\bfeps_-$ for right- and left-handed circular polarization, respectively.
For fixed $\bfzeta_i$ and $\bfeps_\lambda$, but for
unobserved final electronic spin states $m_s$, the triply differential cross section for the emission of bremsstrahlung 
 can be written in the following way \cite{Jaku16},
$$
\frac{d^3\sigma}{d\omega d\Omega_k d\Omega_f}(\bfzeta_i,\bfeps_\lambda)\;=\;\frac{4\pi^2 \omega k_fE_iE_f}{c^5\,k_i\;f_{re}}
$$
\begin{equation}\label{2.2}
\times \; \sum_{m_s=\pm \frac12} \;
\left| \sum_{m_i} a_{m_i} \sum_{\sigma=\pm} f_\sigma\;M_{fi}(\bfeps_\sigma^\ast,m_i,m_s)\right|^2,
\end{equation}
where $\bfk_i, E_i$ and $\bfk_f,E_f$ are momentum and total energy of the electron in its initial and final state, respectively,
$\omega$ is the photon frequency, $f_{re}$ a recoil factor, and $d\Omega_k$ and $d\Omega_f$ are the
solid angles for photon and electron emission, respectively.

We choose a coordinate system where the $z$-axis is taken along the beam direction $\hat{\bfk}_i$. Furthermore, $\bfe_y=\hat{\bfk}_i \times \hat{\bfk}$ and $\bfe_x=\bfe_y \times \hat{\bfk}_i$ where 
$\bfk= k(\sin \theta_k,0,\cos \theta_k)$ is the photon momentum,
such that the reaction ($\bfk_i,\bfk$) plane coincides with the $(x,z)$ plane.
The initial and final electronic states are expanded in terms of partial waves \cite{Ros,PL,Jaku16}, e.g.
$$ \psi_i(\bfzeta_i,\bfr)\;=\;\sum_{m_i=\pm \frac12} a_{m_i}\;\phi_i(\bfr),$$
\begin{equation}\label{2.3}
\phi_i(\bfr)=\sum_{\kappa_i}\sqrt{\frac{2l_i+1}{4\pi}}\;(l_i 0\,\frac12 \,m_i|\,j_i m_i)\;i^{l_i}\;e^{i\delta_{\kappa_i}}\;\psi_{\kappa_im_i}(\bfr),
\end{equation}
where $\psi_{\kappa m}$ is the four-component partial-wave Dirac spinor which is solution to the Dirac equation in the potential generated by the nuclear charge distribution \cite{VJ},
$\delta_\kappa$ is the phase shift and $(\cdot | \cdot )$ a Clebsch-Gordan coefficient \cite{Ed}.
The transition matrix element $M_{fi}$ is defined by
$$M_{fi}(\bfeps_\sigma^\ast,m_i,m_s)\;=\;\sum_{l_fm_l} Y_{l_fm_l}(\hat{\bfk}_f)\;(-i)^{l_f}\;\sum_{j_fm_f} 
e^{i\delta_{\kappa_f}}
$$
\begin{equation}\label{2.4}
\times \;(l_f m_l \,\frac12\,m_s|\,j_f m_f)\;
\int d\bfr\;\psi^+_{\kappa_f m_f}(\bfr)\;\bfalpha \bfeps_\sigma^\ast \;e^{-i\bfks \bfrs}\;\phi_i(\bfr).
\end{equation}
Here $Y_{lm}$ is a spherical harmonic function and $\bfalpha$ a vector of Dirac matrices.
Upon partial-wave expanding $e^{-i\bfks \bfrs}$ and performing the angular integrals one obtains $M_{fi}$ in its most symmetric form,
$$M_{fi}(\bfeps_\sigma^\ast,m_i,m_s)\;=\;i\sqrt{3} \sum_{l_f m_l}Y_{l_f m_l}(\hat{\bfk}_f)\;(-i)^{l_f}
\sum_{j_fm_f} $$
$$
(l_fm_l\frac12\,m_s|\,j_fm_f\!)
\!\sum_{\kappa_i}\!\! \sqrt{2l_i+1}\,i^{l_i}(l_i0\,\frac12\,m_i|\,j_im_i)\,e^{i(\delta_{\kappa_i}\!\!+\delta_{\kappa_{\!f}}\!)}$$
$$
\left\{ \sum_\lambda \sqrt{2\lambda +1} \;\sqrt{\frac{2l_i'+1}{2l_f+1}}\;(-i)^\lambda\;
(l_i'\,0\,\lambda\,0|\,l_f\,0)\;R_{12}(\lambda)\right.
$$
\begin{equation}\label{2.5}
\times \;W_{12}(c_\varrho^{(\sigma)},m_i,l_i',l_f) 
\;-\sum_\lambda \sqrt{2\lambda +1}\;
\sqrt{\frac{2l_i+1}{2l_f'+1}}\;
(-i)^\lambda
\end{equation}
$$ \left.
\times \;
(l_i0\,\lambda \,0|\,l_f'\,0)\;R_{21}(\lambda)\;W_{12}(c_\varrho^{\sigma},m_i,l_i,l_f')\right.\mbox{\LARGE $\}$}
. $$
In this expression, $l=|\kappa+\frac12|\,-\,\frac12$ and $l'=|\kappa - \frac12|\,-\,\frac12\;\;$ (such that $|l'-l|=1$).
Further, $R_{12}$ and $R_{21}$ are the radial integrals,
\begin{equation}\label{2.6}
{ R_{12}(\lambda) \choose R_{21}(\lambda)}\;=\;\int_0^\infty r^2dr\;j_\lambda(kr)\;{g_{\kappa_f}(r) f_{\kappa_i}(r) \choose f_{\kappa_f}(r) g_{\kappa_i}(r) },
\end{equation}
where $j_\lambda$ is a spherical Bessel function and $g_\kappa$ and $f_\kappa$ are, respectively, the large and small components of the radial Dirac function.
From the selection rules of the Clebsch-Gordan coefficients in the curly brackets of (\ref{2.5}), $\lambda$ is restricted to $|l_i'-l_f|\;\leq\lambda\leq \;l_i'+l_f\;$ (step 2) in the first sum, and to $|l_i-l_f'|\;\leq \lambda \leq \;l_i+l_f'\;$ (step 2) 
in the second sum.
The angular part reads
$$W_{12}(c_\varrho^{(\sigma)}, m_i,l_i,l_f)\;=\;\sum_{m_{s_f} m_{s_i}} \sum_{\mu_f \mu_i}\sum_{\mu \varrho} Y_{\lambda \mu}(\hat{\bfk})\;
c_\varrho^{(\sigma)}
$$
\begin{equation}\label{2.7}
(l_f\mu_f\frac12\,m_{s_f}|\,j_fm_f)\;(l_i\mu_i\frac12\,m_{s_i}|\,j_im_i)\;(\frac12\,m_{s_i} 1\,\varrho|\,\frac12\,m_{s_f})
\end{equation}
$$\times \;(l_i\mu_i\lambda\,\mu|\,l_f \mu_f),
$$
where $\bfeps_\sigma^\ast = \sum_{\varrho=0,\pm1} c_\varrho^{(\sigma)}\,\bfe_\varrho\;$ has been expanded in terms of the spherical unit vectors $\bfe_\varrho$ \cite{Ed}.
The expansion coefficients $c_\varrho^{(\sigma)}$ are given by
$$c_\varrho^{(+)}\;=\;\left\{ \begin{array}{cc}
\frac12\;(\cos \theta_k -1), & \varrho =1\\
\\
-\frac12\;(\cos \theta_k +1), & \varrho =-1\\
\\
\sin \theta_k /\sqrt{2}, & \varrho=0
\end{array} \right.$$
\begin{equation}\label{2.8}
c_\varrho^{(-)}\;=\;(-1)^\varrho\;c_{-\varrho}^{(+)}.
\end{equation}

Note that in the numerical evaluation of (\ref{2.5}) with (\ref{2.7}) one has to make use of the fact that the sums over $m_f,\mu_i,\mu_f,\mu,\varrho$ become trivial, since the Clebsch-Gordan coefficients imply
$$
m_f=m_l+m_s,\quad \mu_i=m_i-m_{s_i},\quad \mu_f=m_f-m_{s_f},$$
\begin{equation}\label{2.9}
\mu= \mu_f-\mu_i,\qquad \varrho=m_{s_f}-m_{s_i}.
\end{equation}

\subsection{Polarization correlations}

If the electron spin (unit) vector $\bfzeta_i$ is characterized by the spherical coordinates $(1,\alpha_s,\varphi_s)$, the coefficients $a_{m_i}$ are defined by \cite{Ros}
\begin{equation}\label{2.10}
a_{\frac12}\;=\;\cos \frac{\alpha_s}{2}\;e^{-i\varphi_s/2},\qquad a_{-\frac12}\;=\;\sin \frac{\alpha_s}{2}\;e^{i\varphi_s/2}.
\end{equation}
The basis vectors for a linearly polarized photon are given by $\bfeps_{\lambda_1}=(0,1,0)$ and $\bfeps_{\lambda_2}=(-\cos \theta_k,0,\sin \theta_k)$.
An arbitrary real (linear) polarization vector can likewise be represented in terms of the circular polarization vectors $\bfeps_\pm$ by means of
$$\bfeps_\lambda(\varphi_\lambda)\;=\;
\sin \varphi_\lambda \;\bfeps_{\lambda_1}\;+\;\cos \varphi_\lambda \;\bfeps_{\lambda_2}$$
\begin{equation}\label{2.11}
=\frac{1}{\sqrt{2}}\left(\;e^{-i \varphi_\lambda}\;\bfeps_+^\ast\;+\;e^{i\varphi_\lambda}\;\bfeps_-^\ast \right),\quad \varphi_\lambda \in [0,\pi],
\end{equation}
where $\varphi_\lambda$ is the tilt angle of $\bfeps_\lambda$ out of the reaction plane. 

The polarization correlations $C_{ij0}$ are defined by the following representation of the cross section \cite{T02},
\begin{equation}\label{2.12}
\frac{d^3\sigma}{d\omega d\Omega_k d\Omega_f}(\bfzeta_i,\bfeps_\lambda)=\frac12\left( \frac{d^3\sigma}{d\omega d\Omega_k d\Omega_f}\right)_{\!\!0}\!(1+ \sum_{i,j} \zeta_i \xi_j\;C_{ij0}),
\end{equation}
where $\zeta_i$ and $\xi_j$ describe the electron and photon polarizations, respectively \cite{TP73,PMS,PRT},
while the last zero in the subscript of $C_{ij0}$ relates to the observed unpolarized scattered electron.
The pre\-factor denotes the cross section for unpolarized particles. 
In coplanar geometry (i.e. when $\bfk_f$ lies in the reaction plane),
there are again seven independent polarization correlations $C_{ij0}$ \cite{T02}.
 
Originally, the polarization parameters were termed $P_i$ (Stokes parameters, see, e.g., \cite{OM,MYS}), 
but the $P_i$ agree with the $C_{ij0}$ up to a possible sign.
From (\ref{2.12}) it follows that the $C_{ij0}$, respectively the $P_i$, can be obtained from relative cross section differences (see, e.g. \cite{OM,Jaku10,Jaku12}) which is the method used in experiments (see, e.g. \cite{Na,Ta13}).
The Stokes parameters $P_1,P_2$  for the linear photon polarization  are calculated from
\begin{equation}\label{2.13}
P_i(\alpha_s)\;=\;\frac{d^3\sigma(\bfzeta_i,\bfeps_\lambda(\varphi_\lambda))-d^3\sigma(\bfzeta_i,\bfeps_\lambda(\varphi_\lambda+\frac{\pi}{2}))}{d^3\sigma(\bfzeta_i,\bfeps_\lambda(\varphi_\lambda))+d^3\sigma(\bfzeta_i,\bfeps_\lambda(\varphi_\lambda + \frac{\pi}{2}))},
\end{equation}
where $d^3\sigma$ is defined by the rhs of (\ref{2.2}), dropping the normalization prefactor.
For the Stokes parameter $P_1$, which is independent of $\alpha_s$, one has $\varphi_\lambda = 0$.
However, $\bfzeta_i$ has to be in-plane ($\varphi_s=0)$.
For the Stokes parameter $P_2(\alpha_s)$, one takes 
$\varphi_\lambda = \frac{\pi}{4}$.
When $\bfzeta_i$ is taken along the $z$-axis ($\alpha_s=0$), (\ref{2.13}) yields $P_2(0)=-C_{310}$, while for $\bfzeta_i$ along the $x$-axis ($\alpha_s=90^\circ)$ one obtains $P_2(90^\circ)=C_{110}$.
The circular polarization correlations $P_3(\alpha_s)$ can also be obtained from (\ref{2.13}) upon replacing $\bfeps_\lambda(\varphi_\lambda)$ with $\bfeps_+$ and $\bfeps_\lambda(\varphi_\lambda + \frac{\pi}{2})$ with $\bfeps_-$. Then one has $P_3(0)=C_{320}$ if $\varphi_s=0$ and $\alpha_s=0$, whereas 
  $\varphi_s=0$ and $\alpha_s=90^\circ$ give
$P_3(90^\circ)=-C_{120}$.

On the other hand, the perpendicular spin asymmetry $A$, which does not depend on the photon polarization, is calculated from
\begin{equation}\label{2.14}
A\;=\;\frac{\sum_\lambda d^3\sigma(\bfzeta_i,\bfeps_\lambda)-\sum_\lambda d^3\sigma(-\bfzeta_i,\bfeps_\lambda)}{\sum_\lambda d^3\sigma(\bfzeta_i,\bfeps_\lambda)+\sum_\lambda d^3\sigma(-\bfzeta_i,\bfeps_\lambda)},
\end{equation}
where here, $\bfzeta_i$ has to be perpendicular to the reaction plane $(\varphi_s=-\frac{\pi}{2},\alpha_s=\frac{\pi}{2}),$ and the sum runs over two basis vectors  of the photon polarization.

The last polarization correlation is accessible via
\begin{equation}\label{2.15}
C_{230}\;=\;\frac12 \left( P_1(\bfzeta_i)-P_1(-\bfzeta_i)\right),
\end{equation}
where $P_1(\bfzeta_i)$ is calculated from (\ref{2.13}) with $\varphi_\lambda =0,$ but with an average over the two directions of $\bfzeta_i$ in the denominator.
 $\bfzeta_i$ is defined as in (\ref{2.14}) by $\varphi_s=-\frac{\pi}{2},\alpha_s=\frac{\pi}{2}$ and $-\bfzeta_i$ by $\varphi_s=-\frac{\pi}{2}, \alpha_s=\frac{3\pi}{2}$.

For the calculation of the cross section and the polarization correlations we note that
there are 8 independent amplitudes $M_{fi}$ entering into (\ref{2.2}) since each of the three spin variables can attain two values.
For the denominator of (\ref{2.13}), which is the unpolarized cross section, we obtain from (\ref{2.2}) (with 
the geometry for $P_1$ at $\alpha_s=0$ and 
$f_+=f_-=\frac{1}{\sqrt{2}}$ for $\bfeps_{\lambda_2}$ and $f_+=-f_-=-\frac{i}{\sqrt{2}}$ for $\bfeps_{\lambda_1}$, see (\ref{2.11})),
$$M_{tot}\equiv \sum_{m_s}\left\{ \left| \frac{1}{\sqrt{2}}\;M_{fi}(\bfeps_+^\ast,\frac12,m_s)+\frac{1}{\sqrt{2}}\;M_{fi}(\bfeps_-^\ast,\frac12,m_s)\right|^2 \right.$$
\begin{equation}\label{2.16}
+\;\left. \left| -\frac{i}{\sqrt{2}}\;M_{fi}(\bfeps_+^\ast,\frac12,m_s)+\frac{i}{\sqrt{2}}\;M_{fi}(\bfeps_-^\ast,\frac12,m_s)\right|^2 \right\}
\end{equation}
$$=\sum_{m_s} \left\{ \left| M_{fi}(\bfeps_+^\ast,\frac12,m_s)\right|^2\;+\;\left| M_{fi}(\bfeps_-^\ast,\frac12,m_s)\right|^2 \right\}.
$$
For convenience, guided by \cite{PL}, the following abbreviations are introduced,
$$M_{fi}(\bfeps_\pm^\ast,\frac12,\frac12)\;=\;J_\pm,\qquad M_{fi}(\bfeps_\pm^\ast,-\frac12,\frac12)\;=\;S_\pm,$$
\begin{equation}\label{2.17}
M_{fi}(\bfeps_\pm^\ast,\frac12,-\frac12)\;=\;K_\pm,\qquad M_{fi}(\bfeps_\pm^\ast,-\frac12,-\frac12)\;=\;T_\pm.
\end{equation}
Then, one gets the result
$$M_{tot}=|J_+|^2\;+\;|J_-|^2\;+\;|K_+|^2\;+\;|K_-|^2,$$
\begin{equation}\label{2.18}
P_1\;=\;2 \mbox{ Re } (J_+J_-^\ast +K_+K_-^\ast)/M_{tot}.
\end{equation}
The other polarization correlations are calculated in the same way.
For $P_2(90^\circ)$, one obtains for example,
$$P_2(90^\circ)\;=\;[ \mbox{Im }(J_+J_-^\ast +J_+S_-^\ast +S_+J_-^\ast +S_+S_-^\ast)$$
\begin{equation}\label{2.19}
+\;\mbox{Im } (K_+K_-^\ast +K_+T_-^\ast +T_+K_-^\ast +T_+T_-^\ast)\,]/M_{tot},
\end{equation}
whereas the respective denominator of (\ref{2.13}) leads to
$$M_{tot}\;=\;\frac12\;(|J_+|^2+|S_+|^2+|J_-|^2+|S_-|^2)$$
$$\;+\;\frac12\;(|K_+|^2+|K_-|^2+|T_+|^2+|T_-|^2)$$
\begin{equation}\label{2.20}
+\;\mbox{Re }(J_+S_+^\ast +J_-S_-^\ast)\;+\;\mbox{Re } (K_+T_+^\ast+K_-T_-^\ast).
\end{equation}
From the comparison with $M_{tot}$ from (\ref{2.18}) it follows that the eight amplitudes have to be interrelated.

\subsection{The sum rule}

Guided by the photoeffect studies of \cite{SP,PRT} where time-reversal invariance reduces the 
independent amplitudes  in the cross section to four, we derive a symmetry relation between the $M_{fi}$.
To this aim we must get rid of the sum over $m_s$ which is an incoherent sum. Using the explicit representation
(\ref{2.5}) with (\ref{2.7}) of $M_{fi}$, we can show that
\begin{equation}\label{2.21}
M_{fi}(\bfeps_-^\ast,-m_i,-m_s)=(-1)^{m_i-m_s} M_{fi}(\bfeps_+^\ast,m_i,m_s).
\end{equation}
This is done by reversing for the lhs not only the sign of $m_i$ and $m_s$, but the sign of all other magnetic quantum numbers
appearing in (\ref{2.5}) and (\ref{2.7}) as well. This is possible because these are all summed over.
In order to arrive at the rhs of (\ref{2.21}) the following  symmetry relations are used,
$$(j_1m_1j_2m_2|\,j\,m)\;=\;(-1)^{j_1+j_2-j}(j_1-m_1j_2 -m_2|\,j\,-m),$$
\begin{equation}\label{2.22}
Y_{lm}(\Omega)\;=\;(-1)^m\;Y_{l-m}^\ast(\Omega),
\end{equation}
as well as (\ref{2.8}) for the reversal of the circular polarization.
For eliminating phase factors, the selection rules $l_i'+\lambda+l_f =$ even and $l_i+\lambda +l_f'=$ even, together with (\ref{2.9}), are also applied.
We note that the photon angular function $Y_{\lambda \mu}(\hat{\bfk})$ in (\ref{2.7}) is real because $\bfk$ lies in the $(x,z)$ plane.
On the other hand, the electron angular function has to satisfy
\begin{equation}\label{2.23}
Y_{l_fm_l}^\ast(\hat{\bfk}_f)\;=\;Y_{l_fm_l}(\hat{\bfk}_f)\;e^{-2im_l\varphi}\;\stackrel{!}{=}\;Y_{l_fm_l}(\hat{\bfk}_f),
\end{equation}
which requires a coplanar geometry $(\varphi \in \{0,\pi\})$.
Recall that basically the Clebsch-Gordan algebra has been used for the derivation of (\ref{2.21}).
In particular, no information on the electronic wavefunctions entering into the radial integrals (\ref{2.6}) is necessary.
Also, no restriction of the photon or electron momenta is needed (beyond the requirement that $\bfk_i,\bfk_f$ and $\bfk$ lie in one plane).

From (\ref{2.21}) one has $K_\pm=-S_\mp$ and $T_\pm=J_\mp$, leading to the following simple representations of the polarization correlations,
$$P_1\;=\;C_{030}\;=\;2\mbox{ Re } (J_+J_-^\ast +S_-S_+^\ast)/M_{tot}$$
$$P_2(0)\;=\;-C_{310}\;=\;2\mbox{ Im } (J_+J_-^\ast +S_-S_+^\ast)/M_{tot}$$
$$C_{120}\;=\;2\mbox{ Re }(J_-^\ast S_--J_+S_+^\ast)/M_{tot}$$
$$A\;=\;C_{200}\;=\;2\mbox{ Im }(J_-^\ast S_--J_+S_+^\ast)/M_{tot}$$
$$P_2(90^\circ)\;=\;C_{110}\;=\;2\mbox{ Im }(J_+S_-^\ast -J_-S_+^\ast)/M_{tot}$$
$$C_{230}\;=\;-2\mbox{ Im }(J_+S_-^\ast +J_-S_+^\ast)/M_{tot}$$
\begin{equation}\label{2.24}
C_{320}\;=\;(|J_+|^2+|S_-|^2-|J_-|^2-|S_+|^2)/M_{tot}
\end{equation}
and of the unpolarized cross section,
\begin{equation}\label{2.25}
M_{tot}\;=\;|J_+|^2+|S_-|^2+|J_-|^2+|S_+|^2.
\end{equation}
Thereby, use was made of Re$\;z=$ Re$\;z^\ast$ and Im$\;z=- $ Im$\;z^\ast$.
We note that this functional dependence on the four amplitudes $J_\pm$ and $S_\pm$ is identical to the one for photoionization 
at forward emission \cite{PL} (apart from a possible sign) if $K_\pm$ in \cite{PL} is identified with $S_\pm$ in (\ref{2.24}) and (\ref{2.25}),
and if the correspondence between the polarization correlations in photoionization and bremsstrahlung \cite{PMS} is used.
Profiting from (Re$\,(z))^2+$ (Im$\,(z))^2=|z|^2$
and from Im$\,(z)=(z-z^\ast)/2i$,
 the sum rule (\ref{2.1}), multiplied by $M_{tot}^2$,  is readily verified with the help of (\ref{2.24}) and (\ref{2.25}).

\begin{figure}[!h]
\vspace{-1.0cm}
\includegraphics[width=12cm]{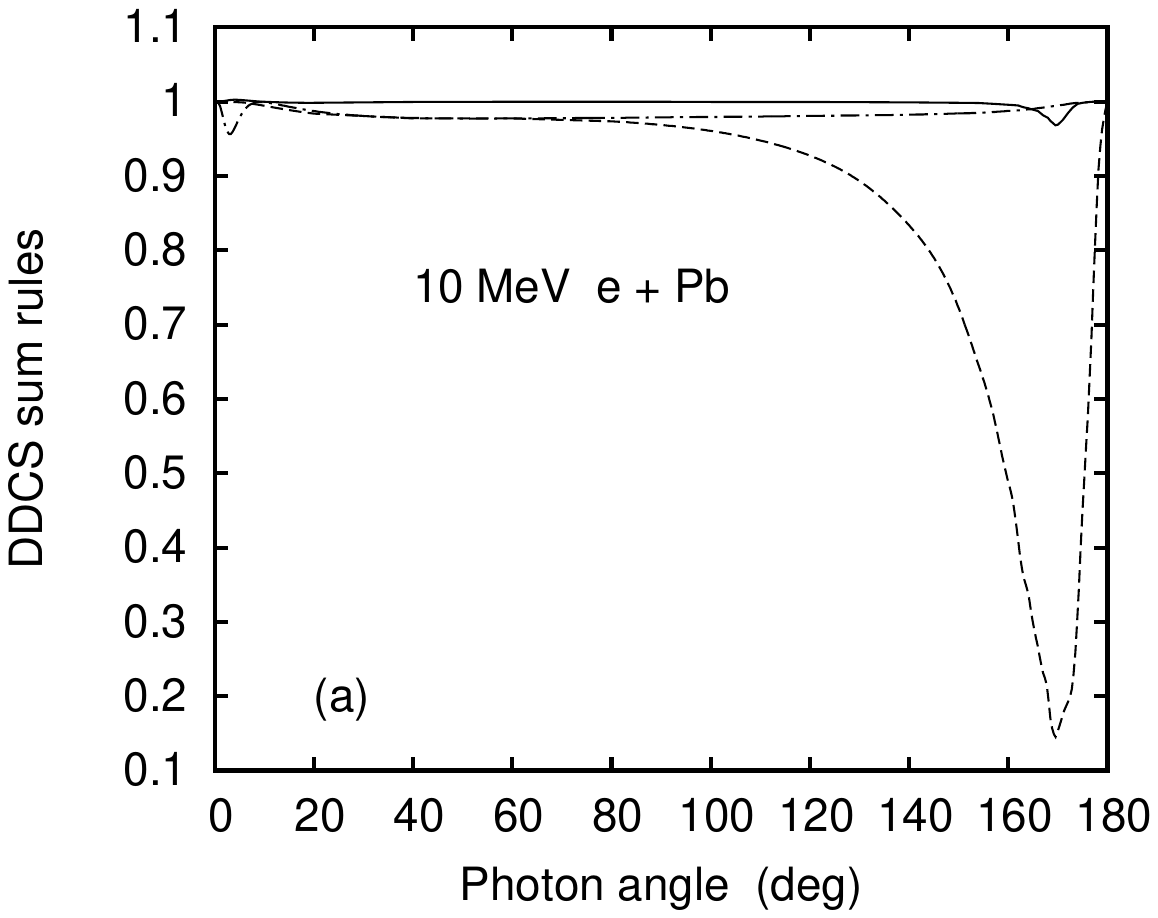}
\includegraphics[width=12cm]{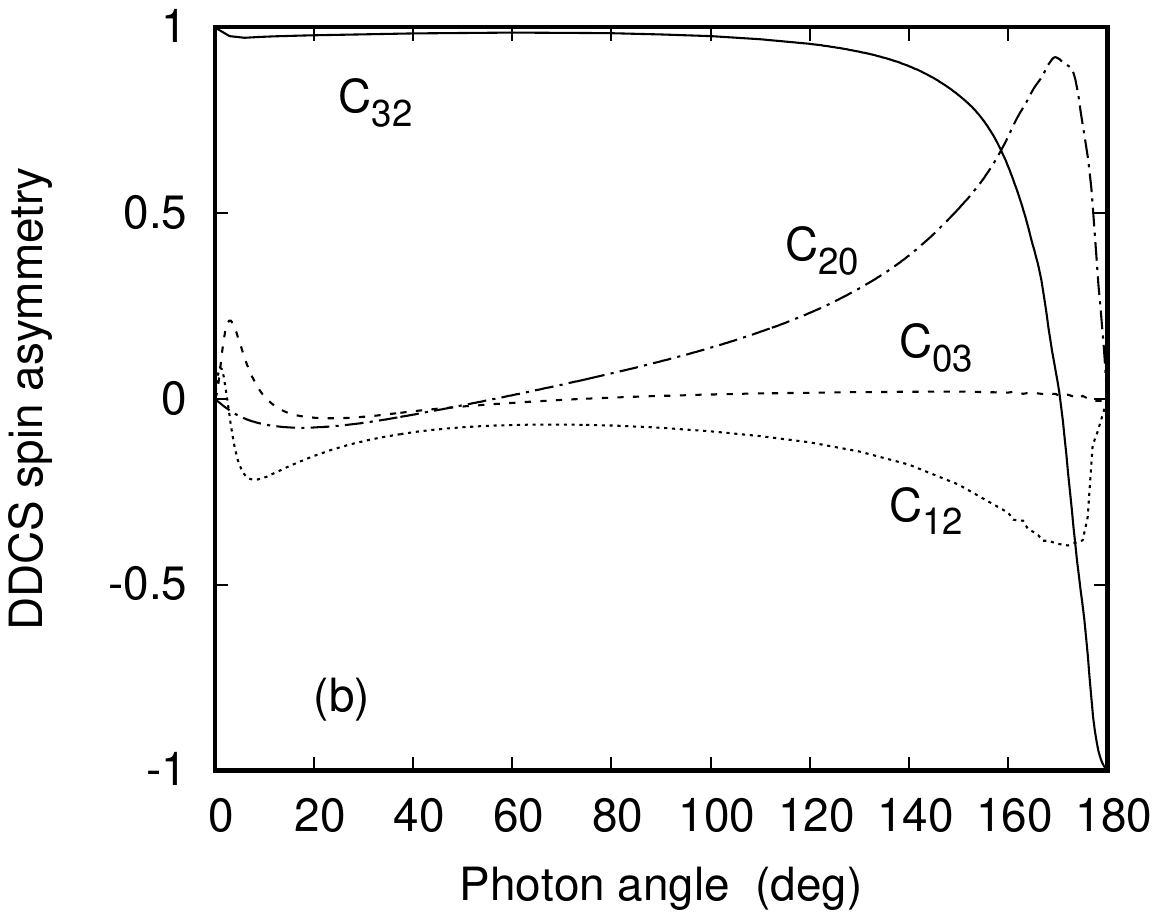}
\caption{
a) Sum rules for the polarization correlations pertaining to the doubly differential cross section for bremsstrahlung from 10 MeV $e+^{208}$Pb ($Z=82$) collisions
at $\omega=9.9$ MeV
as a function of photon emission angle $\theta_k$.
Only $s_{1/2}$-waves are considered for the electronic final state. -----------, exact sum rule (\ref{1.2}); $-\cdot - \cdot -$, sum rule (\ref{1.1}); $-----$, sum rule (\ref{1.3}).\\
b) Spin asymmetries $C_{32}$ (----------), $C_{12}
\;(\cdots\cdots),\;C_{20}\;(-\cdot - \cdot -)$ and $C_{03}\;\;(-----)$,
pertaining to the geometry of Fig.1a, as a function of photon angle $\theta_k$.
Only $s_{1/2}$-waves are considered for the electronic final state.
}
\end{figure}

From the above formalism the results of \cite{PMS} for the polarization correlations pertaining to the doubly differential cross section are easily derived.
The doubly differential cross section is calculated from the same transition matrix element as appears in (\ref{2.4}), but with the basic difference that the partial waves are added incoherently
(see, e.g. \cite{Jaku16}).
Thus, this cross section can still be represented in terms of amplitudes $J_\pm$, $S_\pm,\;K_\pm$ and $T_\pm$ which now depend additionally on the final-state partial wave numbers
$\kappa_f$ and $m_f$.
Only if just a single $\kappa_f$ contributes to this sum, together with two $m_f$ states of opposite sign (which correspond to the two $m_s$ states in (\ref{2.2})),
the symmetry relation (\ref{2.21}) remains valid, from which  the sum rule (\ref{1.2}) follows.

\section{Numerical test of the sum rules}
\setcounter{equation}{0}

The radial Dirac functions entering into the partial-wave representation of the electronic scattering states
are  obtained by solving the Dirac equation with the help of the Fortran code RADIAL \cite{Sal}. The nuclear potential for $^{208}$Pb is generated from the Fourier-Bessel expansion of the nuclear charge density \cite{VJ}. For a nucleus with charge number $Z_T=4$ a pure Coulomb potential is used since for
collision energies of a few MeV nuclear size effects
play no role.
For the evaluation of the radial integrals the complex-plane rotation method \cite{VF,YS} is applied in a refined numerical code \cite{Jaku16}. Recoil, being very small for
our cases of interest, is neglected throughout.
Since only bare nuclei are considered, any screening effects by target electrons are disregarded.

\begin{figure}[!h]
\vspace{-1.0cm}
\includegraphics[width=12cm]{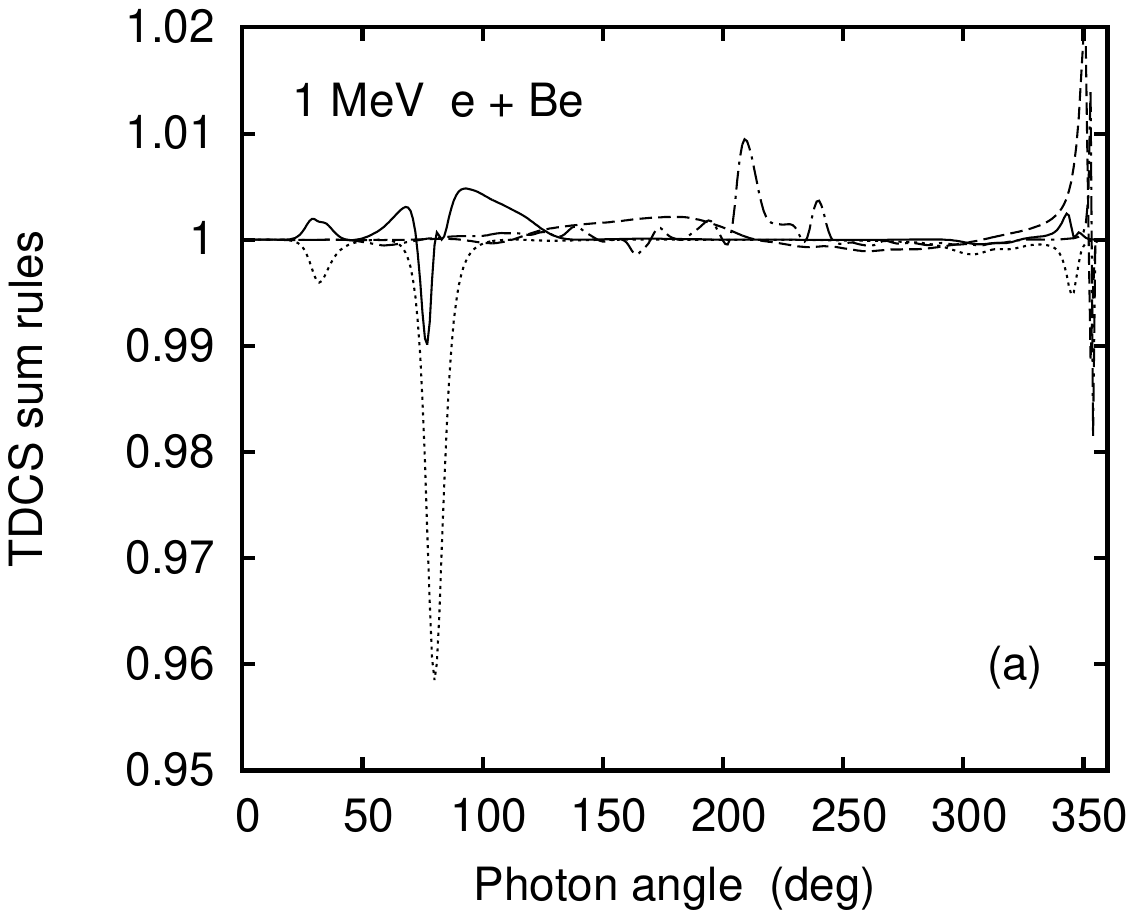}
\includegraphics[width=12cm]{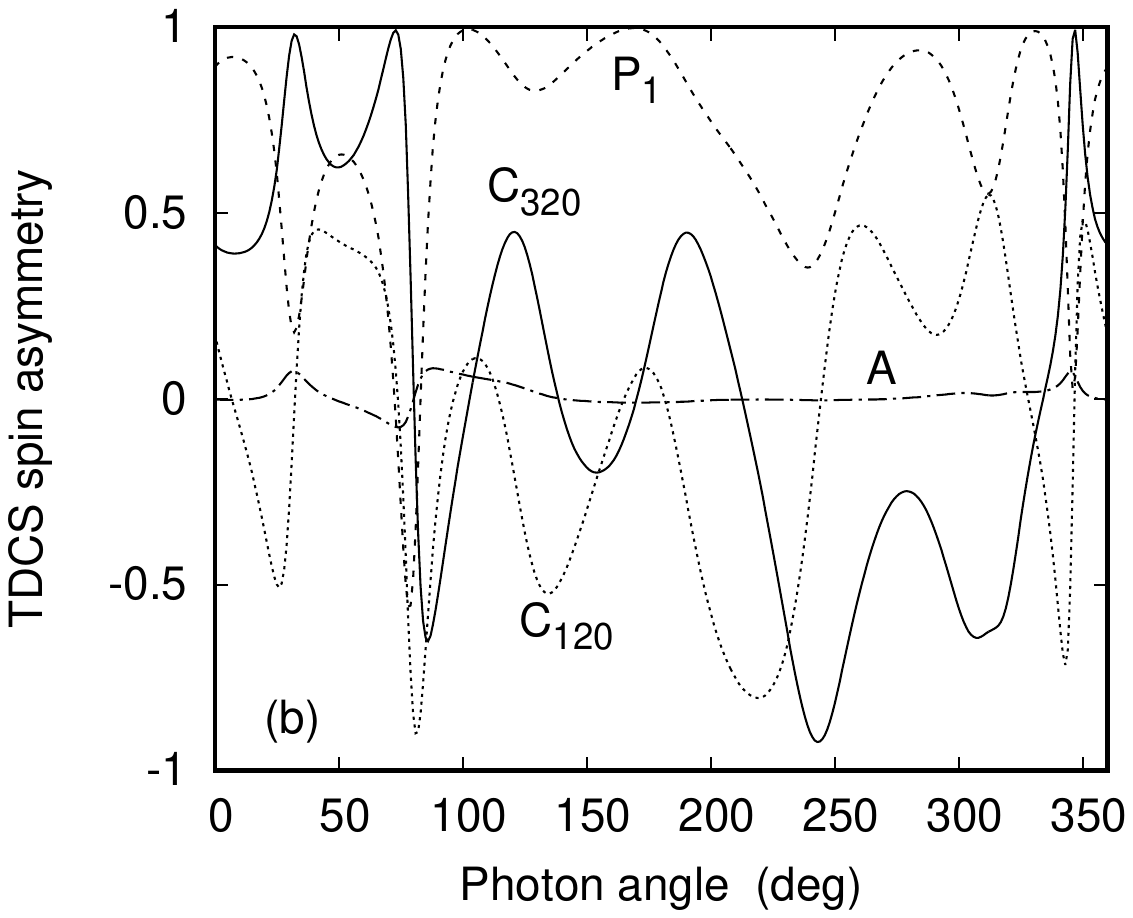}
\caption{
a) Sum rules for the polarization correlations in the elementary process of bremsstrahlung in coplanar geometry ($\varphi =0$) 
from 1 MeV electrons impinging on a Be target ($Z_T=4$)  as a function of photon angle $\theta_k$.
For the sum rule (\ref{2.1}), the final  energy $E_{f,kin}$ is 0.5 MeV and $\vartheta_f=20^\circ$ (--------), 0.1 MeV and $\vartheta_f = 20^\circ \;(-\cdot - \cdot -)$, and 0.1 MeV and $\vartheta_f=90^\circ \;(-----)$.
Also shown is the sum rule  (\ref{3.2}) for 0.5 MeV and $20^\circ \;(\cdots\cdots).$\\
b) Spin asymmetries $C_{320}$ (---------),
$C_{120}\;(\cdots\cdots),\;A\;(-\cdot - \cdot -)$ and $P_1\;(-----)$ for  $Z_T=4,\;E_{i,kin}=1$ MeV, $E_{f,kin}=0.5 $ MeV, $\vartheta_f = 20^\circ$ and $\varphi=0$ as a function of photon angle $\theta_k$.
}
\end{figure}

\subsection{Doubly differential cross section (DDCS)}

By taking $Z_T=4$ and choosing the same collision geometry as in \cite{PMS}
$(E_{i,kin}=E_i-c^2=1$ MeV, $\omega=0.5 $ MeV, restriction to $l_f=0$ final states),
the results from Table III of \cite{PMS} for the seven polarization correlations could be verified,
as well as the validity of the sum rule (\ref{1.2}), within 0.1\%.
However, for a heavy nucleus such as $^{208}$Pb, this sum rule is no longer satisfied to such an accuracy.
This is shown in Fig.1a where a collision energy of 10 MeV and $\omega =9.9$ MeV was chosen.
The largest deviation from unity occurs at backward photon angles $(\theta_k \sim 170^\circ)$ and amounts to 3\%.
This mirrors the accuracy of the partial-wave
approach, the convergence being poorest at the backmost angles.
Results for the sum rule (\ref{1.1}) are also displayed.
They are close to unity, mostly within 1\% except at very small angles $\theta_k$.
However, the sum rule (\ref{1.3}), suggested in \cite{PMS}, is strongly violated for such a  high nuclear charge.
This can be explained by the angular distributions of the spin asymmetries $C_{20}$ and $C_{03}$ (see Fig.1b)
which differ strongly from each other in the backward regime.
While $C_{03}$ remains close to zero, $C_{20}$, increasing proportional to $Z_T/c$, approaches its maximum value 1 near $170^\circ$ 
at a collision energy around 10 MeV.
If all partial waves of the final state are included, the polarization correlations change (see \cite{Jaku16} for $C_{20}$),
resulting in a strong violation of the sum rule (\ref{1.2}).
Note that the approximate sum rules (\ref{1.1}) and (\ref{1.3}) imply no restrictions on the final state.
However, at very large impact energies, a condition for their validity, it is mostly the $s_{1/2}$ waves which contribute to the emission of hard photons.

\subsection{Triply diferential cross section (TDCS)}

Since the polarization correlations for the elementary process of bremsstrahlung obey the sum rule (\ref{2.1}) for arbitrary final electronic states,
we have considered in the plots of this section all final partial waves 
which are necessary for the convergence of the partial-wave expansion. For a final kinetic energy $E_{f,kin}=0.5$ MeV, about twenty partial waves are included, while for $E_{f,kin}=0.1$ MeV, ten
 are sufficient ($|\kappa_f| \leq 11$, respectively $|\kappa_f| \leq 5)$.

Apart from examining (\ref{2.1}), we also consider the approximate sum rules corresponding to (\ref{1.1}),
\begin{equation}\label{3.1}
C_{320}^2+C_{120}^2+A^2\;\approx\;1,
\end{equation}
and to (\ref{1.3}),
\begin{equation}\label{3.2}
C_{320}^2+C_{120}^2+P_1^2\;\approx\;1.
\end{equation}

In our first example we show in Fig.2a the angular dependence of the sum rules for $Z_T=4$ and a collision energy of 1 MeV.
Except near $\theta_k=350^\circ$ the sum rule (\ref{2.1}) is, irrespective of the electronic final momentum, satisfied within 1\%,
 which is less accurate than the doubly differential case treated in \cite{PMS}.
Also the sum rule (\ref{3.2}) holds for most angles within 1\%.
For this low-$Z_T$ target, $A$ is very small while the first Stokes parameter $P_1$ oscillates strongly with a large amplitude (Fig.2b).
Consequently, the replacement of $P_1$ by $A$ in this sum rule 
induces very large deviations from unity. Of course, the resulting sum rule (\ref{3.1}) is not expected to hold at low $Z_T$.

\begin{figure}[!h]
\vspace{-1.0cm}
\includegraphics[width=12cm]{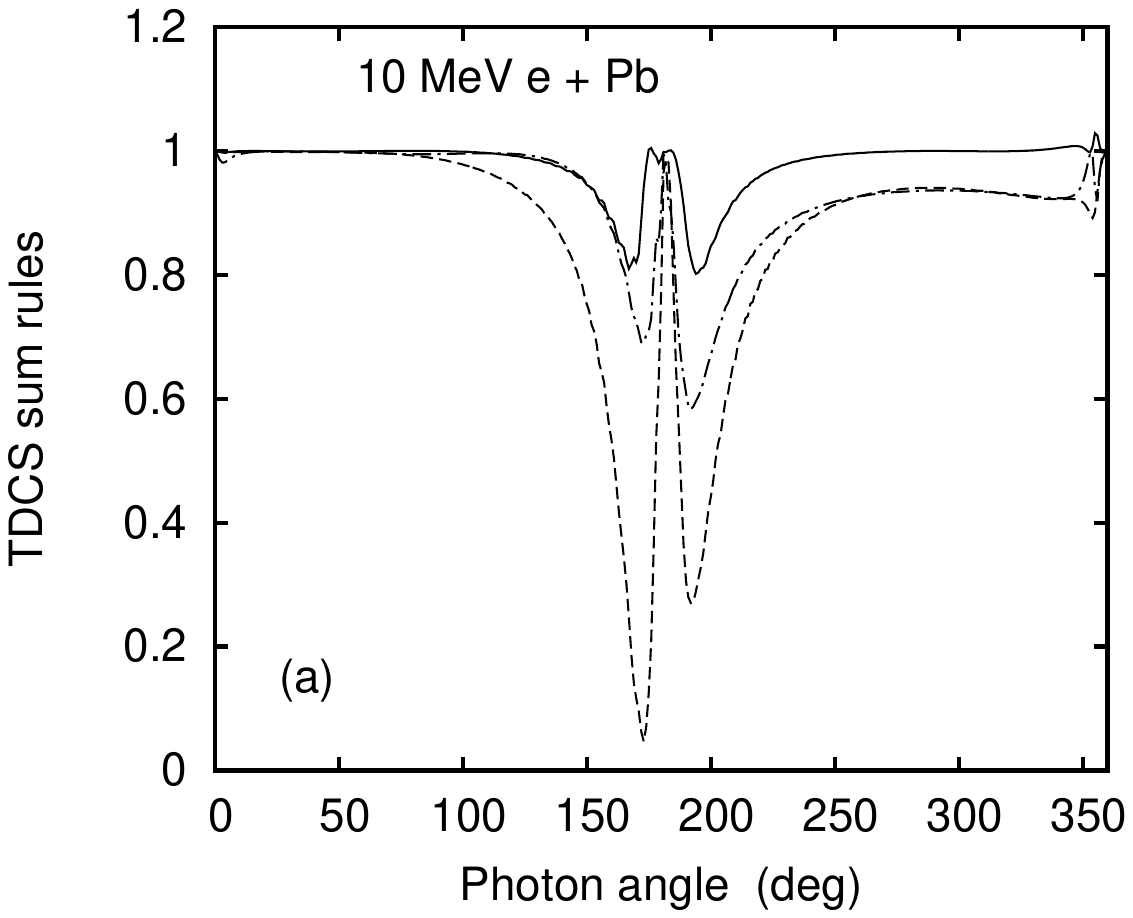}
\includegraphics[width=12cm]{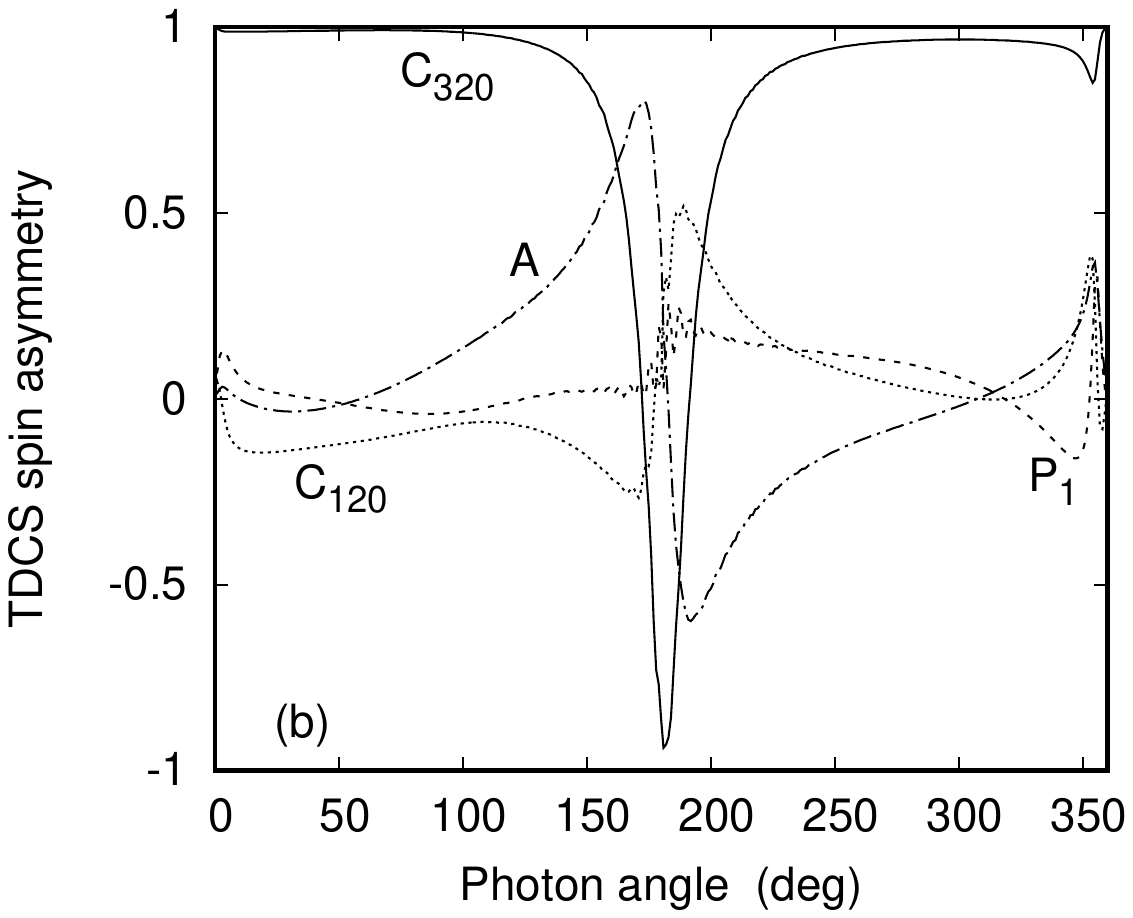}
\caption{
Elementary process of bremsstrahlung
 from 10 MeV $^{208}$Pb$(e,e')$ collisions
at $\vartheta_f=90^\circ$ for $\omega = 9.9$ MeV as a function of photon angle $\theta_k\;\,(\varphi=0)$.\\
a) Sum rules for the polarization correlations:
----------, sum rule (\ref{2.1}); $-\cdot - \cdot -, $ sum rule (\ref{3.1}); $-----$, sum rule (\ref{3.2}).\\
b) Spin asymmetries $C_{320}$ (---------), $C_{120} (\cdots \cdots),\;\;A\;(-\cdot - \cdot -)$ and $P_1 \;(----)$.
}
\end{figure}

The heavy nucleus $^{208}$Pb is treated in Fig.3.
While in the case of an unobserved electron the inclusive sum rule (\ref{1.2}) for the spin asymmetries 
is quite well satisfied,
this is no longer true for (\ref{2.1}) where the scattered electron is observed in coincidence with the bremsstrahlung photon.
For the same collision parameters as in Fig.1 ($E_{i,kin}=10$ MeV, $E_{f,kin}=0.1$ MeV),
the deviations from unity amount up to 20\% in an extended $\theta_k$-region around $180^\circ$ (Fig.3a).
Neither of the approximate sum rules (\ref{3.1}) or (\ref{3.2}) are valid for this nucleus.
Matters do not improve when the scattering angle is varied.
However, the sum rules (\ref{2.1}) and (\ref{3.1})
are at most photon angles the better satisfied, the higher the collision energy and the less energetic the scattered electron.

Fig.3b displays the corresponding polarization correlations.
It is seen that for $\theta_k < 180^\circ$ their angular dependencies closely resemble those shown in Fig.1b.
However, in particular for the spin asymmetries $C_{120}$ and $P_1$, convergence problems in an angular regime around $180^\circ$
manifest themselves in terms of (unphysical) wigg\-les.
These inaccuracies are the cause of the strong violation of the sum rules (\ref{2.1}) and (\ref{3.1}) for such angles.

\vspace*{0.5cm}

\section{Conclusion}

Using the relativistic partial-wave formalism, an exact sum rule for the polarization correlations pertaining to the elementary process of bremsstrahlung has been derived.
Although this sum rule is shown to be valid in coplanar geometry for any collision parameters such as nuclear charge as well as energy or emission angle of the participating particles, a numerical proof fails when the nucleus is heavy.
Thus the deviations from unity may serve as a measure of the accuracy of the numerical calculations.
However, one has to keep in mind 
that the
polarization correlations, which are related to cross section differences,
 are determined with a much poorer accuracy 
that the
 cross sections themselves.

Finally we want to point out that the present theory, although considering finite nuclear size effects, disregards magnetic scattering
and the dynamical recoil which are important for electron scattering from nuclei with spin when the
collision energy exceeds a few tens of MeV \cite{BL}.
An additional (incoherent) summation over the (unobserved) final nuclear spin states will of course spoil the sum rule.

\vspace{1cm}

\noindent{\bf ACKNOWLEDGMENT}

It is a pleasure to thank R.H.Pratt for stimulating this work and for many helpful discussions.



\vspace{1cm}

\end{document}